\documentclass[aps,prb,preprint,superscriptaddress,amsmath,amssymb,floatfix]{revtex4-1}
\usepackage{lmodern}
\usepackage[utf8]{inputenc}
\usepackage[T1]{fontenc}
\usepackage{amsmath}
\usepackage{amssymb}
\usepackage{textcomp}
\usepackage{graphics}
\usepackage{graphicx}

\usepackage{xcolor}

\newcommand{\uam}{Departamento de F\'{i}sica Te\'{o}rica de la Materia
                  Condensada, Universidad Aut\'{o}noma de Madrid,
                  E-28049 Madrid, Spain}
\newcommand{\uamfmc}{Departamento de F\'{i}sica de la Materia Condensada, Universidad
                     Aut\'{o}noma de Madrid, E-28049 Madrid, Spain}
\newcommand{\icmm}{Instituto de Ciencia de Materiales de Madrid (ICMM),
                   CSIC, c/ Sor Juana Ines de la Cruz 3, E-28049 Madrid, Spain}
\newcommand{\ifimac}{Condensed Matter Physics Center (IFIMAC),
                     Universidad Aut\'{o}noma de Madrid, E-28049 Madrid, Spain}

\begin{document}  

\title{Tailoring the thermal expansion of graphene via controlled defect creation}

\author{Guillermo L\'{o}pez-Pol\'{i}n}
\affiliation{\uamfmc}
\altaffiliation{G. L.-P. and M. O. contributed equally to this work.}

\author{Maria Ortega} \affiliation{\uam} \altaffiliation{G. L.-P. and M. O. contributed equally to this work.}

\author{J. G. Vilhena} \affiliation{\uam} \affiliation{\icmm} 

\author{Irene Alda} \affiliation{\uamfmc}

\author{J. Gomez-Herrero} \affiliation{\uamfmc} \affiliation{\ifimac} 

\author{Pedro A. Serena} \affiliation{\icmm}

\author{C. Gomez-Navarro} \email{cristina.gomez@uam.es} \affiliation{\uamfmc} \affiliation{\ifimac} 

\author{Rub\'{e}n P\'{e}rez} \email{ruben.perez@uam.es} \affiliation{\uam} \affiliation{\ifimac} 




\begin{abstract}
Contrary to most materials, graphene exhibits a negative thermal expansion coefficient (TEC), i.e it contracts when heated.
This contraction is due to the thermal excitation of low energy out--of--plane vibration modes. These flexural modes
have been reported to govern the electronic transport and the mechanical response of suspended graphene.
%
In this work, we systematically investigate the influence of defects in the TEC of suspended graphene membranes. Controlled introduction of low densities of mono-vacancies reduces the graphene TEC, up to one order of magnitude for a defect density of $5\times10^{12}$~cm$^{-2}$. Our molecular dynamics simulations reproduce the observed trend and show that TEC reduction is due to the suppression of out--of--plane fluctuations caused by the strain fields created by mono-vacancies in their surrounding areas. These results highlight the key role of defects in the properties of ``real-life'' graphene, and pave the way for future proposals of electronic and mechanical defect engineering.

See Supplementary information at:

http://uam.es/departamentos/ciencias/fismateriac/pdfs/julio/Graphene\_TEC\_with\_SI\_CondMat.pdf
\end{abstract}



\maketitle

Graphene exhibits intrinsic out--of--plane thermal fluctuations that have dramatic
effects on its  conformation~\cite{Meyer2007},  electronic~\cite{Laitinen2014,Bolotin2008} and thermal transport properties~\cite{Seol2010}, and on its elastic response~\cite{Katsnelson2013}.
One of the most prominent consequences of the presence of these flexural modes is the graphene large negative thermal expansion
coefficient (TEC)~\cite{Bao2009,Singh2010,Yoon2011,Mounet2005,Zakharchenko2009,
Zakharchenko2010,deAndres2012,Gao2014,Jiang2015}.
The amplitude of these very soft acoustic ZA phonons rapidly increases with temperature and translates into an actual contraction of the material. 

%
%
The thermodynamical theory of membranes predicts that out--of--plane thermal should renormalize the elastic constants of graphene, making it softer as the amplitude of the fluctuations grows~\cite{Nelson2004,Katsnelson2013}. López-Polin et al.~\cite{Lopez-Polin2014} reported an increase in the Young’s modulus of graphene for a dilute density of single--atom vacancies. In order to address this counterintuitive result, these authors suggested, without further proof, that the induced defects ironed the thermal fluctuations out, unveiling the bare (non-renormalized) Young’s modulus of graphene. This scenario implies that, not only the Young’s modulus, but all the elastic constants should be affected by the presence of defects. 

%
Graphene TEC is an ideal candidate to explore the interplay between intrinsic thermal vibrations and defects, and to confirm or reject the conjecture made in ref.~\cite{Lopez-Polin2014}. Moreover, this interaction is key to understand the physics underlying many temperature effects in “real-life” graphene, from the behavior of nanoresonators~\cite{Mathew2016} to the diffusion of water nanodroplets on graphene~\cite{Tocci2014}.
However, measuring graphene TEC is a challenging task since conventional experimental techniques designed for bulk materials cannot
be applied to such thin membranes, and the  anchoring of one--atom--thick membranes brings out technological difficulties~\cite{Garza2014}.Therefore only few experimental works are available in literature~\cite{Bao2009,Yoon2011}.
Here, we showed that the graphene TEC can be significantly reduced by the controlled introduction of low densities of single-atom (mono-) vacancies. Our novel approach measures the TEC by the determination of the stress induced in a suspended
graphene sheet due to the mismatch with the substrate TEC during annealing--cooling cycles.  This stress was determined from
nanoindentations performed with the tip of an atomic force microscopy (AFM) on monolayered graphene drums.
Our molecular dynamics (MD) simulations reproduced the observed trend, and revealed that TEC reduction is due to the quenching of out-of-plane fluctuations caused by the strain fields created by mono-vacancies. 

\newpage

\textbf{Results}

Experiments were performed on samples prepared by mechanical exfoliation of natural graphite on Si$O_{2}$/Si substrates with predefined circular wells with diameter of 1--2~$\mu$m (see Methods).
Only monolayered graphene was selected for this study.
Suspended circular membranes were tested by indenting the tip at the center of the clamped area with an AFM probe.
This situation can be modeled as a circular membrane with a central point load where the force versus indentation curves behave as~\cite{Begley2004}:
\begin{equation} \label{eq: indentation force}
F(\delta) = \pi\sigma_0\delta + \frac{E_{2D}}{a^2}\delta^{3}
\end{equation}
where $F$ is the loading force, $\delta$ is the indentation at the central point, $a$ is the drumhead radius, $E_{2D}$ is the
two-dimensional elastic modulus of the membrane, and $\sigma_0$ is the stress of the membrane (both in N/m). Our $F(\delta)$ curves
fitted correctly eq.~\ref{eq: indentation force} (see Supplementary Fig.~1a) 
and our measured values in more than 10 drumheads for $E_{2D}$ (350$\pm$30 N/m) and $\sigma_0$ (0.2$\pm$0.1 N/m) are in agreement
with previous works~\cite{Lee2008,Garcia-Sanchez2008}.The initial $\sigma_0$ reflects a small prestress
accumulated in the sheet during the sample preparation procedure.

\begin{figure}[t]
\includegraphics[width=0.65\columnwidth]{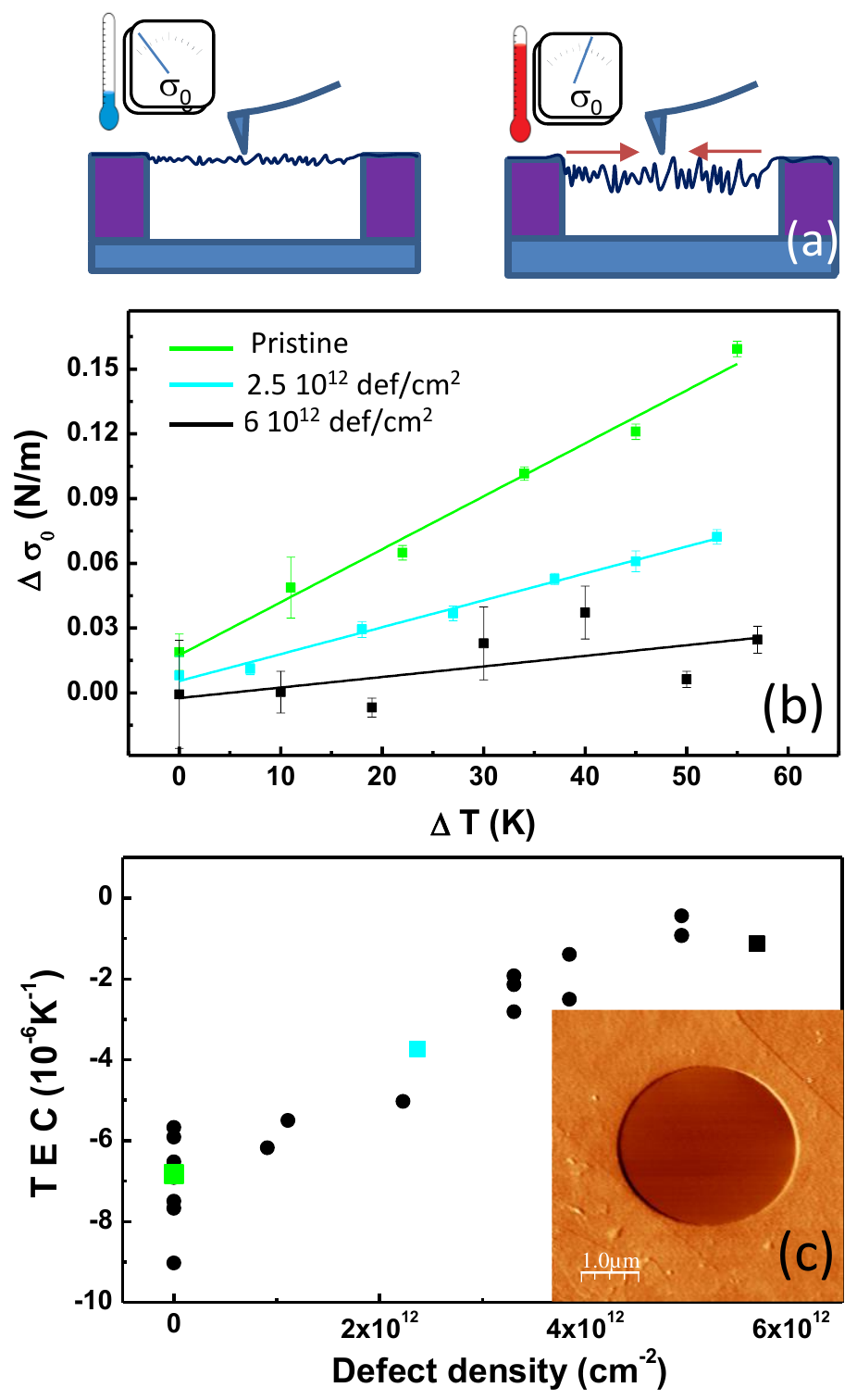}
\caption{
\textbf{Experimental measurements of graphene TEC.}
\textbf{(a)} Sketch of the experimental set-up. \textbf{(b)}
Plot of measured stress as a function of sample temperature. Each color corresponds to a drumhead with
a different defect density. Green: pristine. Blue: $2.5 \times 10^{12}$  $cm^{-2}$.
Black: $6.0 \times 10^{12}$ $cm^{-2}$. \textbf{(c)} Experimentally measured TEC of different
membranes as a function of the induced defect density. Inset: AFM image of a
representative graphene drumhead.}
\label{fig:EXP}
\end{figure}

Towards thermo--mechanical characterization of graphene suspended layers, sample temperature was varied between
10--75 {\textcelsius} and consecutive $F(\delta)$ curves were acquired at intermediate temperatures. Upon annealing,
graphene tends to contract while the Si$O_{2}$ substrate shows negligible expansion~\cite{Yoon2011}.
This difference in behavior leads to an effective increase in the stress of the suspended graphene area
that can be measured through the value of $\sigma_0 (T)$ (see Fig.~\ref{fig:EXP}a and Supplementary Fig.~1b).
The use of the linear coefficient in eqn.~\ref{eq: indentation force} is indeed an experimental novelty since so far that equation was used to determine the Young's modulus of 2D materials through the coefficient of the cubic term.
Green dots in Fig.~\ref{fig:EXP}b represent our measured $\sigma_0$ as a function of sample temperature
for a representative pristine graphene drumhead. As expected, it displays an increasing stress with temperature.
The two dimensional TEC ($\alpha_{2D}$) of graphene can be directly calculated from this graph using:
\begin{equation} \label{eq: EXP_TEC}
\alpha_{2D}=-\left( \frac{\partial \epsilon}{\partial T} \right)_{V}
= -\frac{1}{E_{2D}} \left( \frac{\partial \sigma_0}{\partial T} \right)_{V},
\end{equation}
where $\epsilon$ is the strain induced by temperature.
Our measured mean value for the room temperature TEC in 7 pristine membranes, ~$(-7\pm1)\times 10^{-6}$~K$^{-1}$,
agrees with previous experimental reports~\cite{Bao2009,Yoon2011} and supports the validity of our technique for the
characterization of thermo-mechanical properties of suspended 2D materials.

In order to measure the variation of the graphene TEC in the presence of defects,
a controlled density of point defects was then introduced in the membranes by irradiating the samples
with a known dose of Ar$^+$  with incoming energy of 140 eV in high vacuum. As reported in previous works by our group and others
this technique allows the creation of controlled densities of carbon mono-vacancies~\cite{Lopez-Polin2014,Ugeda2010} :
we carefully characterized the induced defect type and density by Raman spectroscopy and Scanning Tunneling Microscopy (STM) in
ambient conditions as described in ref.~\onlinecite{Lopez-Polin2014} and Section SI1 in the Supplementary Information: Both the ratio of the intensity of the I$_D$ and I$_D’$ peaks~\cite{Eckman2012,Cancado2011} in the
Raman spectra (Supplementary Fig.~2), and the characteristic $ \sqrt{3} \times \sqrt{3}$ pattern observed in the atomically resolved STM images~\cite{Ugeda2010} (Supplementary Fig.~3)
pointed towards clean single vacancies without sp$^3$ hybridization. The  I$_{D}$/I$_G$ relation was used to determine the vacancy density.
Consecutive low dose irradiations enable systematic study as a function of vacancy density

Mechanical testing by indentation experiments was performed after each ionic dose with the same AFM probe and in the
same conditions described above. The  $F(\delta)$ curves in defective membranes
displayed also excellent fitting to eq.~\ref{eq: indentation force} allowing
accurate determination of $\sigma_0$ (see Supplementary Fig.~4). Fig.~\ref{fig:EXP}b illustrates our results for
pristine and defective membranes. The stress of irradiated membranes do not increase as much as that
of pristine membranes with increasing temperature i.e. they exhibit a lower TEC in
absolute value (hereafter referred as |TEC|).
Systematic measurements in more than 10 graphene drumheads are represented in Fig.~\ref{fig:EXP}c.
These results illustrate our main experimental finding: the high negative value of the TEC of graphene decreases
with the introduction of vacancy-like defects, and 
approaches to zero ($\sim -1\times 10^{-6}$~K$^{-1}$) for a defect density of ~$6\times10^{12}$ $cm^{-2}$
(corresponding to a mean distance between vacancies of $\sim 4$~nm).

\begin{figure}[th]
\centering
\includegraphics[width=0.9\columnwidth]{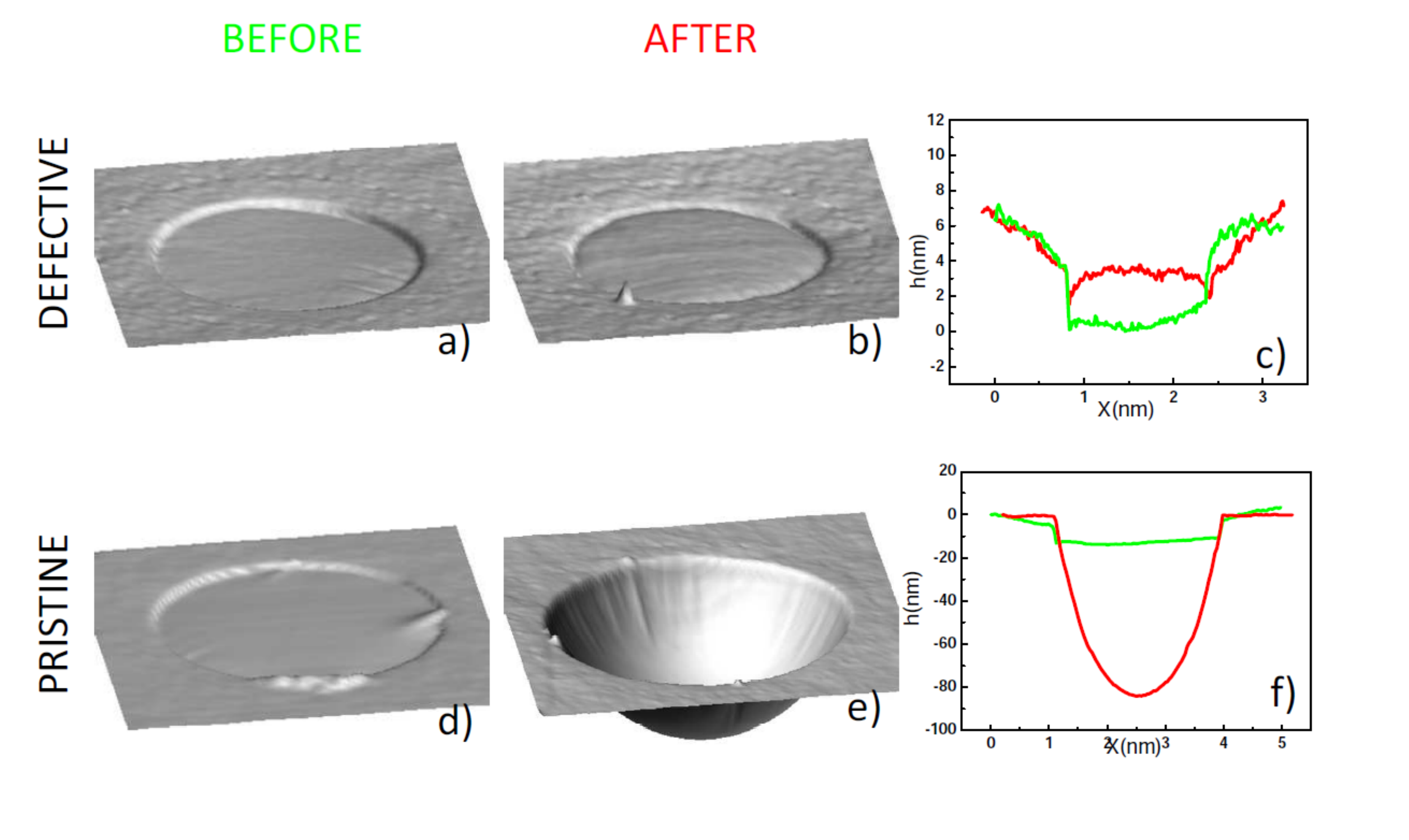}
\caption{
\textbf{AFM images on drumheads before and after an annealing-cooling cycle from 300 to 650 K.}
(\textbf{a}) and (\textbf{b}) panels correspond to a  defective drumhead (defect density $6\times 10^{12}$
cm$^{-2}$), while (\textbf{d}) and (\textbf{e}) are measured on  a pristine graphene drumhead. Panels (\textbf{c}) and (\textbf{f}) are topographic profiles along the great arc of the drumheads before (green) and after (red) the annealing-cooling cycle.}
\label{fig:EXP2}
\end{figure}

As an additional experiment
to test our results through a more established approach,
we performed AFM imaging of defective and pristine graphene
drumheads before and after annealing-cooling cycles. Suspended graphene has been reported
to undergo dramatic morphological changes under such cycles~\cite{Bao2009}. These changes comprise the
appearance of periodic ripples, and the bending of graphene membranes towards the substrate.
Both features are caused by the thermal stress induced in the membrane by the TEC mismatch
between graphene and the 
SiO$_2$ substrate, that has a negligible TEC.
Our images of pristine graphene before and after an annealing-cooling cycle from 300 to 650 K showed these morphological changes. Representative images are portrayed in Figs.~\ref{fig:EXP2}d and \ref{fig:EXP2}e, where periodic ripples in the radial direction and
substantial buckling towards the substrate can be observed. Based on these images, the TEC of pristine graphene can be estimated according to:
$\alpha_{2D}=\frac{L_f}{ L_i  \Delta T}$,
where $L_f$ and $L_i$ are the final and initial arc of the membrane.  The value derived from this alternative methodology ($6.5\times 10^{-6}$~K$^{-1}$) agrees well with that
derived above and in previous reports~\cite{Bao2009}. We performed parallel AFM imaging before and after annealing--cooling
cycles in membranes with a defect density of $6\times 10^{12}$ cm$^{-2}$. In this case, we observe neither buckling nor periodic ripples (Figs.~\ref{fig:EXP2}a and \ref{fig:EXP2}b), indicating that the TEC mismatch
between the substrate and the graphene membrane approaches to zero i.e. a strongly reduced |TEC| in defective
graphene, further supporting the results presented above.


\begin{figure}[t]
\includegraphics[width=0.65\columnwidth]{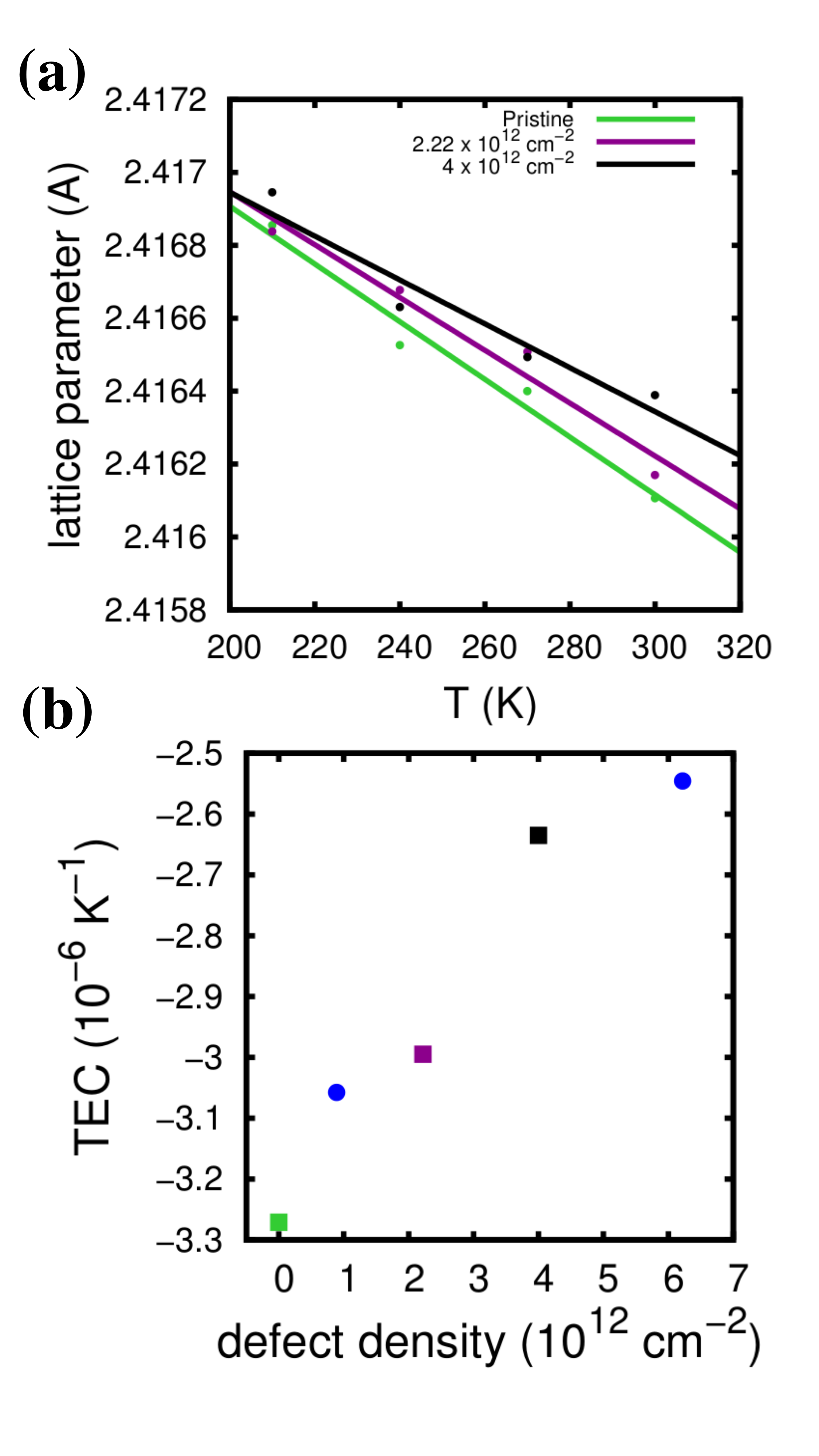}
\caption{
\textbf{Theoretical determination of the TEC from NPT simulations.}
\textbf{(a)} Lattice parameter versus temperature for pristine
(green) and defective ($4\times 10^{12}$~cm$^{-2}$) graphene (black).
The TEC is determined by the slope of the corresponding linear fits (solid lines).
\textbf{(b)} TEC as a function of the defect density. Data for $C_{def} = 4\times 10^{12}$~cm$^{-2}$
is an average of four different defect configurations (see Supplementary Figs.~7b and 11).
}
\label{fig:Theo_TEC}
\end{figure}

As a first step to understand our measurements, we performed constant pressure
(NPT) MD simulations to determine the lattice parameter versus temperature for
pristine and defective graphene (see Methods for further details on the simulations).
Long (40~ns) simulations were performed in $\sim 15\times 15$ nm$^2$ graphene
slabs. Defects were created by single atom removal. Fig.~\ref{fig:Theo_TEC}a
displays the results of these simulations for four different temperatures for
pristine graphene (green) and two membranes with defect concentrations
$2.22\times 10^{12}$~cm$^{-2}$ (purple circles) and $4\times 10^{12}$~cm$^{-2}$ (black circles).
We can quantify the TEC using the linear fits shown in Fig.~\ref{fig:Theo_TEC}a
and the standard TEC formula, $\alpha_{2D}= \frac{1}{l_{0}}
\left( \frac{\partial l}{\partial T} \right)_{P} $.
Our TEC for pristine graphene, $(-3.27\times 10^{-6}$ $K^{-1})$, is in excellent
agreement with previous {\em ab initio}~\cite{Mounet2005} and
classical--potential~\cite{Zakharchenko2009,Gao2014} theoretical studies. More
importantly, our results for the defective systems
(e.g. $(-2.63\times 10^{-6}$ $K^{-1}$ for  $4\times 10^{12}$~cm$^{-2}$ ) show that the introduction of defects leads to
a reduction of the |TEC|, in agreement with our experimental observations.
Our results for different defect concentrations (Fig.~\ref{fig:Theo_TEC}b and Supplementary Fig.~7) confirm the trend found in the
experiments: the |TEC| decreases rapidly as the concentration of defect
increases, and tends to saturate for concentrations beyond
$4\times 10^{12}$~cm$^{-2}$. These results are robust with respect to the defect
distribution and the number of sampling temperatures (see Supplementary Fig.~8).


\begin{figure}[t]

\includegraphics[width=0.60\columnwidth]{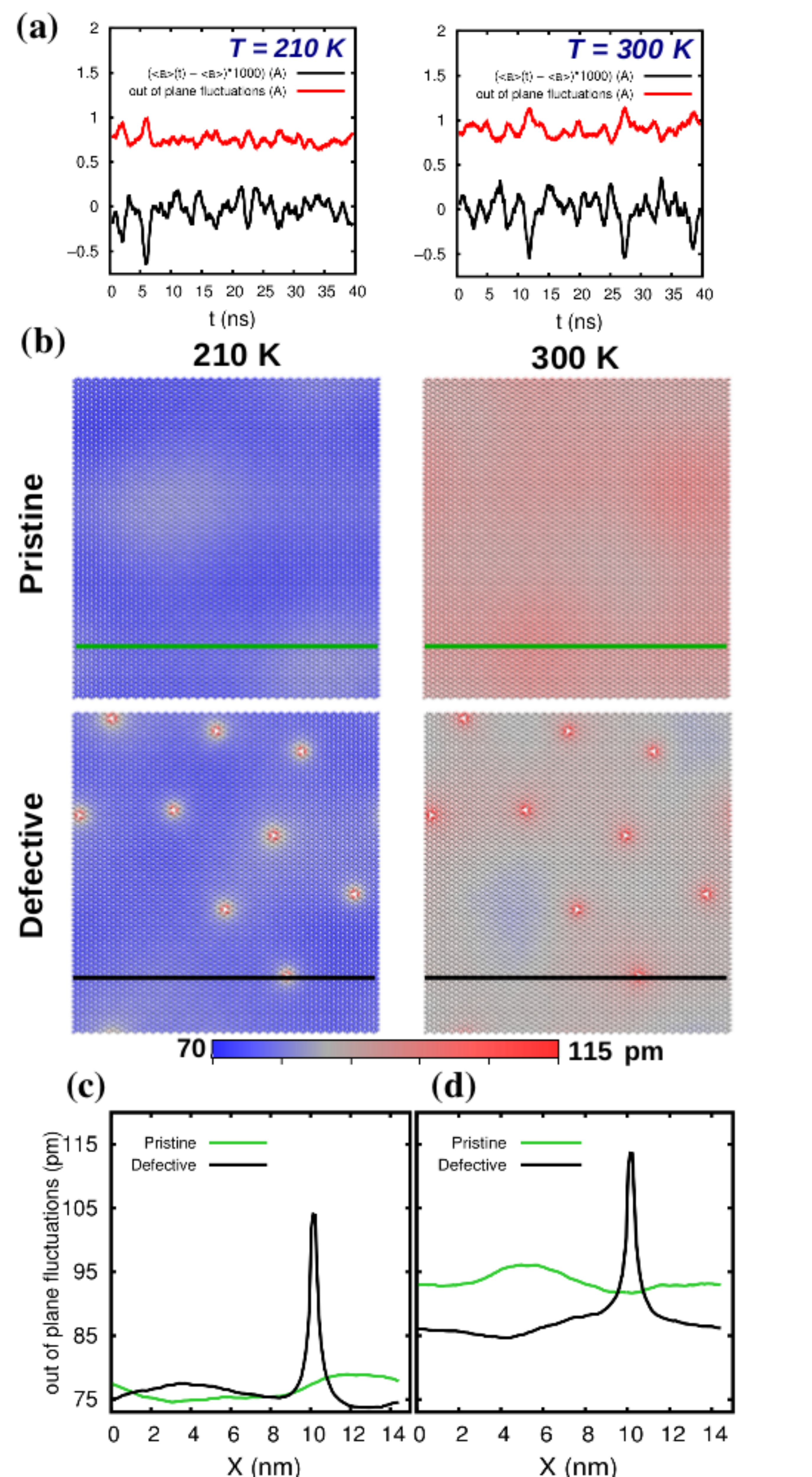}
\caption{
\textbf{Analisis of out-of-plane fluctuations in NPT simulations.}
\textbf{(a)}Time evolution of the spatially--averaged amplitude of the out--of--plane fluctuations $h(t)$ (red),
computed as described in the Methods section, and the fluctuations of the lattice parameter (black) for the simulations at $T = 210$~K and at $T = 300$~K.
\textbf{(b)} RMS amplitude of the out--of--plane fluctuation per atom, $\langle h (\bar{r}_i) \rangle$,
(average of the $z$--coordinate of atom $i$ over the last 20~ns)  for pristine and defective graphene at two temperatures ($T= 210, 300$~K).
\textbf{(c)} Profile of $\langle h (\bar{r}_i) \rangle$ along the green (black) line in panel (c) for pristine (defective) graphene at $T= 210$~K.
\textbf{(d)} Same as (c) for $T=300$~K.
}
\label{fig:Theo_Fluc}
\end{figure}

\bigskip

\textbf{Discussion}

In order to understand the microscopic mechanism behind the TEC reduction,
we explore the changes induced in the out--of--plane fluctuations by the combined effect
of temperature and the presence of defects. Fig.~\ref{fig:Theo_Fluc}a clearly shows
the expected anticorrelation in the time evolution of the lattice parameter and
the spatially averaged root--mean-square (RMS) amplitude for out--of--plane oscillations, $h(t)$ (see Methods).
More insight can be gained looking at the real space distribution (atom by atom)
of the time averaged RMS amplitude of the out--of--plane fluctuations, $\langle
h (\bar{r}_i) \rangle$ (with $\bar{r}_i$ the $xy$ in--plane position of atom
$i$, see Methods), displayed in Fig.~\ref{fig:Theo_Fluc}b for two different
temperatures and in the presence of defects.
These 2D maps capture two important features: First, for pristine graphene, the
dominant red color at 300~K, in contrast to the blue color at 210~K, reflects
that temperature is responsible for the increase in the amplitude of
out--of--plane fluctuations for each atom, leading to an actual in--plane
contraction of the material.
Second, defective graphene exhibits overall smaller fluctuations than pristine
graphene. Although the atoms that are first neighbors of the defects have
larger oscillation amplitudes, fluctuations have been drastically reduced on
the rest of the graphene sheet, as shown by the dominance of grey color in the
bottom left panel of Fig.~\ref{fig:Theo_Fluc}b. This effect can be quantified
using the line profiles shown in Fig.~\ref{fig:Theo_Fluc}c--d. Summarizing, our
simulations show that the TEC reduction in the presence of defects is induced by
the reduction of out--of--plane fluctuations.

At this stage, we speculate that the quenching of out--of--plane fluctuations is linked with an additional in--plane stress induced in the graphene sheet by the defects.
The presence of a uniform tensile strain is known to reduce the out--of--plane fluctuations~\cite{Roldan2011}.
We explored this idea with constant volume (NVT) MD simulations, that closely resemble the methodology used in our experiment, where graphene edges are clamped.
NVT simulations were performed in the same range of temperatures and defect concentration used in the NPT case,
using the lattice parameter obtained for pristine graphene at 210~K. Fig.~\ref{fig:Theo_Stress}a
shows the converged thermal stress $\langle \sigma \rangle$ (see Supplementary Fig.~9) for pristine and defective
($4\times 10^{12}$~cm$^{-2}$) graphene as a function of temperature.
As in the experiments, the thermal stress of pristine graphene increases much faster
than the defective one. This behavior, also observed for other defect densities (see
Supplementary Fig.~10), allows us to independently confirm that a decrease on the thermal
strain indeed translates into a reduction of the TEC.

\begin{figure}[t]
\includegraphics[width=0.55\columnwidth]{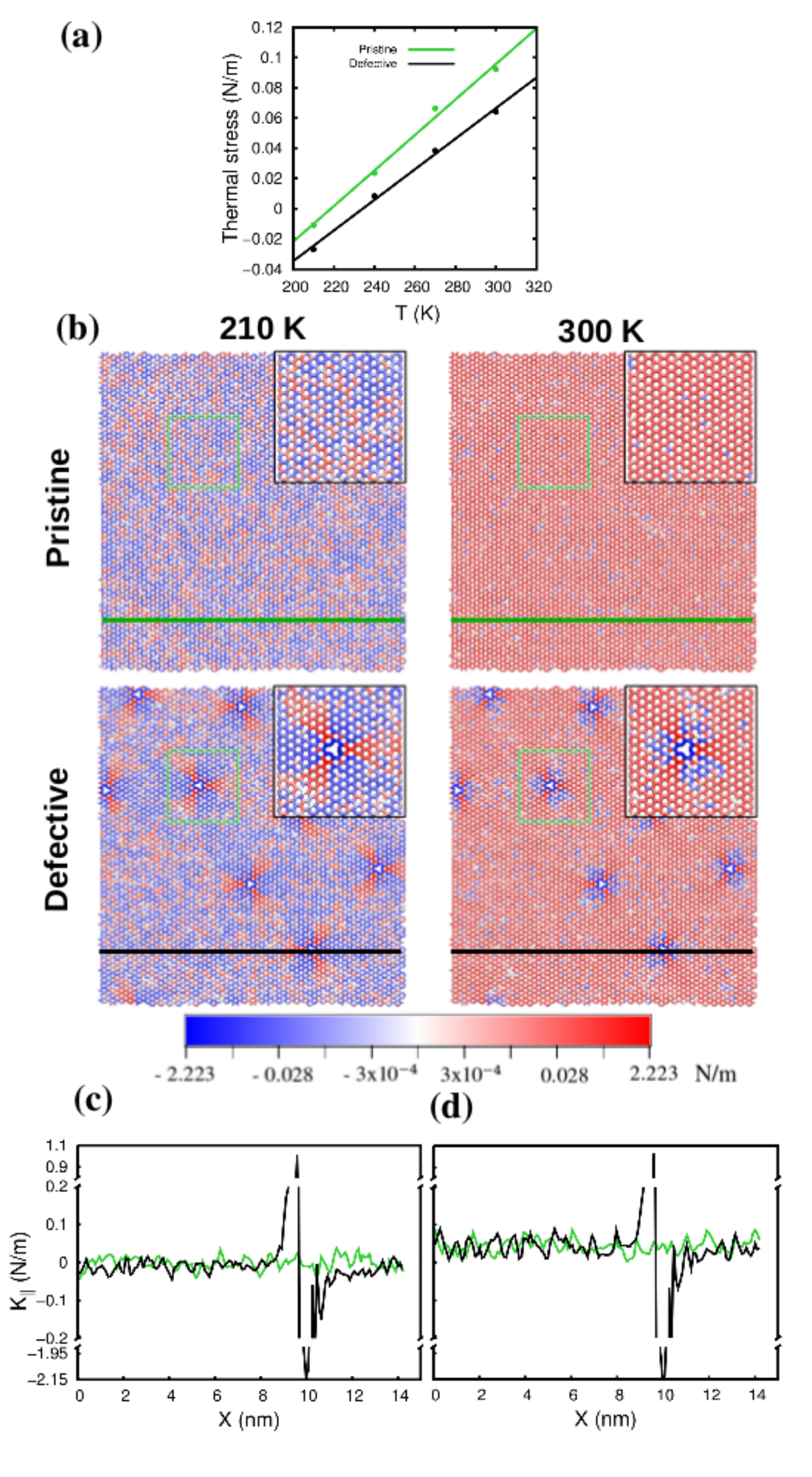}
\caption{
\textbf{Theoretical determination of the stress from NVT simulations.}
\textbf{(a)} Thermal stress versus temperature for  pristine and defective
($4\times 10^{12}$~cm$^{-2}$) graphene using the graphene lattice parameter at $T=
210$~K
\textbf{(b)} Sum of the in--plane diagonal components of the $i$--atom stress tensor
(normalized by the surface area per atom $A_{at} = 0.02529$ nm$^2$)
$ \langle K_{\parallel} (\bar{r}_i) \rangle = \langle (K_{xx} (\bar{r}_i) + K_{yy} (\bar{r}_i))/2 A_{at} \rangle$ for pristine and
defective graphene at $T= 210, 300$~K. Time averages in (a) and (b) include the last 20 ns of the simulations.
\textbf{(c)} Profile for $\langle K_{\parallel} (\bar{r}_i) \rangle$ along the green (black) line in panel (b) for pristine (defective) graphene at $T= 210$~K.
\textbf{(d)}Same as (c) for $T=300$~K. Notice the logarithmic scale used in panels (b)--(d).
}
\label{fig:Theo_Stress}
\end{figure}

In our classical simulations, we are able to decompose the total stress into a sum of local contributions for each of the carbon atoms (see Methods).
Fig.~\ref{fig:Theo_Stress}b presents the time average of the sum of the in--plane diagonal components of the $i$--atom stress
tensor $\langle K_{\parallel}(\bar{r}_{i}) \rangle =  \langle (K_{xx}(\bar{r}_{i}) + K_{yy}(\bar{r}_{i}))/2 A_{at} \rangle$
as a 2D stress map for both pristine and defective graphene at two different temperatures.
In--plane compressive (positive) stress is represented in red, while blue corresponds to in--plane tensile
(negative) stress, that would tend to expand locally the graphene sheet.
For pristine graphene at 210~K,  we have an even distribution of very small regions with low positive and negative stress
that cancel out. This result is consistent with the fact that the simulation box size is fixed to the equilibrium value of pristine graphene at 210~K.
When temperature is raised to 300~K, the 2D stress map turns red, revealing an homogeneous increase of the in--plane
compressive stress in the system (see Fig.~\ref{fig:Theo_Stress}c--d) as observed experimentally. This means
that, if fixed boundary conditions were relaxed, graphene would shrink in--plane.


The 2D stress map for defective graphene at 210~K (see Fig.~\ref{fig:Theo_Stress}b)
looks similar to the one of the pristine case  in areas far from the defects (e.g., the left bottom corner).
However, around defects, we observe a three-fold symmetry stress distribution, where areas of tensile (blue) and compressive
(red) stress alternate. Overall, the map shows a predominance of blue areas with tensile stress. If fixed boundary conditions are released,
this stress would favor an in-plane expansion that naturally leads to a larger equilibrium lattice parameter for the defective case,
in agreement with our NPT results (Fig.~\ref{fig:Theo_TEC}a).
At 300K, we observe an increase of the in--plane compressive stress similar to the one found in the pristine case.
However, at variance with the latter, small patches of tensile stress can still be found around the defects.
These patches are responsible for a reduction of the global compressive stress in the system,  that would lead,
if boundary conditions are removed, to a smaller in--plane contraction compared to the pristine case.

The maximum tensile stresses observed around defects for both temperatures (see Figs~\ref{fig:Theo_Stress}c and~\ref{fig:Theo_Stress}d) are $\sim 2.15$~N/m, that corresponds to a local strain of ~0.65~\%. Theoretical calculations based on the self-consistent screening approximation and atomistic MonteCarlo simulations~\cite{Roldan2011}, predict a very strong attenuation of the out of plane thermal fluctuations in stiff membranes, as graphene, with the presence of rather weak uniform strains ($\sim 0.4$ \%).
Strain fields less than 1\% are enough to suppress all the anharmonic effects.
This strong dependence with the strain provides the fundamental link between our results from the NPT and NVT simulations. Defects support short--wavelength normal vibrations, but the local strain that they induce contributes to the quenching of the long--wavelength thermal fluctuations that are mainly responsible for the in--plane graphene contraction. The remaining fluctuations lead to the reduced negative TEC measured in our experiments and simulations in the presence of defects.
Thus, the stress distribution induced by the defects, in particular, the presence of areas of tensile stress,
is responsible for the |TEC| reduction. 


In spite of the remarkable agreement in experimental and simulated trends, there are quantitative discrepancies regarding the TEC for pristine graphene,
where our theoretical results predict a TEC almost two times smaller than our experiments. This discrepancy has been already pointed
out in previous experimental works~\cite{Bao2009,Yoon2011}.
According to recent MD simulations~\cite{Gao2014}, |TEC| increases with system size for square $L \times L$ unit cells
in the range $L\sim10-100$~nm. This behavior can be understood in terms of the effective cutoff introduced by the finite
system size in the long-wavelength out--of--plane vibration modes that are responsible for the negative TEC of graphene.
This interpretation is consistent with the fact that experimental and theoretical values for the TEC in defective
systems are much closer:  For a defect density of $4\times 10^{12}$~cm$^{-2}$,
the average TEC obtained in the experiments ($\sim -2.4\times 10^{-6}$ K$^{-1}$)
compares very well with our theoretical result ($\sim -2.63\times 10^{-6}$ K$^{-1}$).
According to our simulations (see Fig.~\ref{fig:Theo_Fluc}b), the introduction of defects changes the TEC by
reducing the average out--of--plane fluctuations. The suppression of these normal modes reduces the dependence on
the system size and should bring a better theory--experiment agreement as we increase the defect concentration.
This is precisely what we have found in our study, confirming our conclusions regarding the role of defects in tuning the TEC.
In summary, we have measured the TEC of graphene using a novel approach that can be extended to other 2D materials. We have shown that the TEC can be tuned through the controlled introduction of vacancy-like defects, reducing it to almost zero for a modest defect concentration (corresponding to a mean distance between vacancies of $\sim 4$~nm). Our long MD simulations reproduce this trend, validate the experimental approach, and provide a comprehensive understanding of this phenomenon in terms of the quenching of out--of--plane fluctuations induced by the patches of tensile strain introduced by the defects.
Although the textbook physics of pristine graphene has attracted much attention, our work proves that the unavoidable presence of small defect densities in real--life graphene has profound implications. Our insight provides a guide for future proposals of either electronic or thermal defect engineering~\cite{Haskins2011}. Along this line, we suggest controlled creation of low defect densities to reduce the TEC mismatch between graphene and substrates that cause undesired strains in devices.
Our results have also to be considered in the interpretation of previous reports on the interplay of defects and mechanical properties. In particular, they constitute the first evidence supporting the conjecture of graphene stiffening by quenching of long--wavelength phonons by monovacancies~\cite{Lopez-Polin2014} (see Supplementary Fig.~5), paving the way for the tailoring of the mechanical properties of graphene and other 2D materials where out--of --plane thermal fluctuations are relevant.


\bigskip

\textbf{Methods}

\textbf{Experiments.}
Substrates with circular wells were obtained in Si/SiO$_2$ patterned with optical
lithography and reactive ion etching.  Graphene monolayers were first identified by
optical microscopy and then corroborated by Raman spectroscopy as described in
ref.~\cite{Ferrari2006}.
Raman spectra were performed using a confocal microscope in ambient conditions with laser wavelength 532nm and power  and 0.7mW.
This ensures no damage of the samples. AFM images and indentation curves were acquired with a Nanotec commercial microscope and WSxM software package~\cite{Horcas2007}.
We used silicon AFM probes from Nanosensors with a nominal force constant of 3 N/m, each probe was individually calibrated using the Sader method~\cite{Sader1999}.
Images were obtained in non-contact dynamic mode in order to minimize surface-tip interaction. Indentation curves were performed at the center of graphene drumheads
with an approaching/retracting speed of 90nm/s.  Defect density was monitored by in-situ measuring the sample ionic current during irradiation.
Defect type and density was then ex-situ probed by Raman spectroscopy
and atomically resolved Scanning Tunneling Microscopy
as discussed in ref.~\cite{Lopez-Polin2014} and in section SI1 of the Supplementary Information.

\bigskip

\textbf{Molecular Dynamics Simulations}

\textbf{Interatomic potentials and simulation details.}
%
%
Molecular dynamics (MD) simulations were performed with the LAMMPS software suite~\cite{Plimpton1995} and the reactive
force field AIREBO~\cite{Stuart2000,Brenner2002} to describe the interatomic interactions. AIREBO, the well known second generation of
the reactive empirical bond order (REBO) potential, describes a wide range of mechanical~\cite{Kvashnin2015} and thermal
properties~\cite{Gao2014, Mosterio2014} of pristine~\cite{Gao2014,Mosterio2014,Kvashnin2015} as well as defective~\cite{Mosterio2014,Kvashnin2015}
graphene. In particular, the torsional term of this force field provides a good description of  the low energy out--of--plane phonon modes of
graphene ({\em ZA,ZO}), which are crucial to reproduce the negative TEC of graphene and its change with temperature.

We have modeled a large $(63 a \times 35 \sqrt{3} a) \sim 15\times 15$~nm$^2$ unit cell ($a$ is the graphene lattice parameter)
with different concentrations and distributions of monovacancy defects (see Supplementary Fig.~11) at different temperatures. The unit cell for the pristine case includes 8820 atoms.
We use periodic boundary conditions (PBC) with the corresponding 2D unit cell (for $x$ and $y$ directions) and a large vacuum of 40 nm in the $z$ direction).
In all simulations, Newton's equations of motion have been integrated with the velocity--Verlet integrator with a time step
$\Delta t = 1$~fs. A total simulation time of 40~ns have been used in order to ensure a proper sampling of the thermal fluctuations and
to reach the high accuracy on the lattice parameter ($\sim10^{-5}$\AA) needed to determine changes in the TEC (see Supplementary Fig.~6) and in the stress (see Supplementary Fig.~9.)
For NPT simulations, the temperature and pressure of each simulation are kept constant using the Nose-Hoover thermostat and barostat
as implemented in LAMMPS. The PBCs together with the pressure restrain ($P=0$~atm) allow the box size ($l_x\times\l_y$) to be relaxed to the equilibrium at each temperature value.
For NVT simulations, we use the same protocol but we keep the box size constant (constant volume) using the lattice parameter of pristine graphene at $T=210$~K.
The lower temperatures (210--300~K) chosen for the theoretical analysis compared to the experiment (283--348~K) 
help to achieve the precision needed to determine the changes in the lattice parameter with shorter simulation times.
%

\textbf{Spatial and time averages}

\textbf{(1) Lattice parameter}

The lattice parameters plotted in Figure~2a are determined from the time average (hereafter denoted $\langle \cdots \rangle$)
of the instantaneous value, $a(t)$, for each time step over the last 20 ns, [$t_i=20$~ns  , $t_f=40$~ns], of the corresponding NPT simulation:
\begin{equation}
\langle a(t) \rangle = \frac{1}{\sqrt{N_f}} \sum_{t_i}^{t_f} a(t);  \;\;\;\;  a(t)=\sqrt{\frac{l_{x}\cdot l_{y}}{63\cdot 35\sqrt{3}} }.
\end{equation}
where $a(t)$ is determined from the  instantaneous dimensions $l_{x}(t)$ and $l_{y}(t) $ of the unit cell. $N_f = (t_f -t_i) / \Delta t$ is
the number of time steps that are included in the average. For details on the convergence of $\langle a(t) \rangle$ (see Supplementary Fig.~6).

\textbf{(2) Spatially averaged out--of--plane fluctuation $h(t)$}

We have evaluated  for each time step, $t$, the spatial average, $z(t)$, and standard deviation (RMS), $h(t)$, of the $z$ coordinate of the $N$ atoms of the membrane:
\begin{equation}
h(t)= \frac{1}{\sqrt{N}} \left[ \sum_{i=1}^{N}(z_i(t)- z(t))^2 \right]^{1/2} \!\!\!\!\!\! ;  \;\;\;\; z(t) = \frac{1}{N}\sum_{i =1}^{N} z_i(t)
\end{equation}
$h(t)$ represents the amplitude of the out--of--plane fluctuations of the membrane.

\textbf{(3) Time-averaged out-of-plane fluctuation per atom (2D maps) $\langle h (\bar{r}_i) \rangle$}

The 2D maps in Figure~3b display $\langle h (\bar{r}_i) \rangle = \langle (z_i(t)- z_i)^2 \rangle^{1/2}$,
with $z_i(t)$ the instantaneous $z$ coordinate for atom $i$ in the unit cell, and $z_i = \langle z_i(t) \rangle $, its time average.

\textbf{(4) Time--averaged Thermal stress and per--atom stress tensor}

The thermal stress $\sigma$ is linked with the per-atom-stress tensor~\cite{Gao2014}:
\begin{equation}
\sigma = \frac{1}{A} \sum_{i=1}^{N} (K_{xx}^i + K_{yy}^i + K_{zz}^i) = \frac{1}{N} \sum_{i=1}^{N}  \frac{1}{A_{at}} (K_{xx}^i + K_{yy}^i + K_{zz}^i),
\end{equation}
where $K_{aa}^i = K_{aa}(\bar{r}_i)$ are the diagonal components ($a = x, y, z$) of the per--atom stress tensor  for atom $i$; $A = N A_{at}$
is the total area of the unit cell, with  $A_{at}$ the area per atom, and $N$ the number of atoms.

We have calculated $\sigma$ in our NVT simulations as a time average over the last 20 ns of the instantaneous pressure $P(t)$, using the relations:
\begin{equation}
\sigma = -\frac{3V}{A} \langle P(t) \rangle; \;\;\;\; P(t) = - \frac{1}{3 V} \sum_{i=1}^{N} \left[ K_{xx}^i (t)+ K_{yy}^i (t) +K_{zz}^i (t) \right].
\end{equation}
For details on the convergence of $\langle \sigma \rangle$ see Supplementary Fig.~9.

The 2D stress maps shown in Figure~4b correspond to the time average (last 20 ns) of the in--plane stress per atom (normalized by $A_{at}$) $\langle K_{\parallel}(\bar{r}_i) \rangle$:
\begin{equation}
\langle K_{\parallel}(\bar{r}_i) \rangle =   \frac{1}{2 A_{at}} \langle K_{xx}(\bar{r}_i,t) +K_{yy}(\bar{r}_i,t) \rangle
\end{equation}
Our simulations confirmed that the contribution of the out--of--plane stress  $\langle K_{zz}(\bar{r}_i) \rangle$ is approximately 10 times smaller than the in--plane contribution.


\bigskip

\textbf{Acknowledgements}

We acknowledge financial support from the Spanish MINECO (projects CSD2010-00024,
MAT2011-023627, MAT2013-46753-C2-2-P, MAT2014-54484-P, MDM-2014-0377, and FIS2015-69295-C3) and from the MAD2D-CM Program (project S2013/MIT-3007) of the Comunidad de Madrid.
Computational resources provided by the  Red Espa\~{n}ola de Supercomputaci\'{o}n
(RES) and the Extremadura Research Centre for Advanced Technologies (CETA-CIEMAT).
We acknowledge fruitful discussions with Paco Guinea, Pablo San Jose, Lucia
Rodrigo, Pablo Pou and Lars Pastewka.

\bigskip

\textbf{Author contributions}

G. L.-P., I. A., J. G.-H. and C. G.-N. performed the experiments and the corresponding data analysis.
M. O., J. G. V., P. A. S., and R.P. carried out the molecular dynamics simulations and and analyzed the
theoretical results. All of the authors contributed to the discussion and the writing of the paper.

\bigskip

\textbf{Competing financial interests:} The authors declare no competing financial interests.



\end{document}   